\documentclass[graybox]{svmult}

\usepackage{type1cm}        
\usepackage{makeidx}         
\usepackage{graphicx}        
\usepackage{multicol}        
\usepackage[bottom]{footmisc}

\usepackage{newtxtext}       %
\usepackage[varvw]{newtxmath}       

\usepackage{algorithm}
\usepackage{algorithmic}
\usepackage{amsmath}
\usepackage{tikz}
\usepackage{pgfplots}
\usepgfplotslibrary{groupplots}
\usetikzlibrary{calc}
\pgfplotsset{compat=1.18}

\definecolor{bblue}{HTML}{4F81BD}
\definecolor{rred}{HTML}{C0504D}
\definecolor{ggreen}{HTML}{9BBB59}
\definecolor{ppurple}{HTML}{9F4C7C}

\makeindex             

\begin{document}

\title*{Optimal Transport Metric Learning for Feature Alignment in Partially Supervised Segmentation}
\titlerunning{Optimal Transport Metric Learning in Partially Supervised Segmentation}
\author{Dakini Mallam Garba\orcidID{0009-0007-5198-1506} and\\Salim Abdou Daoura\orcidID{0009-0005-7128-552X}}
\institute{Dakini Mallam Garba \at ML Collective, \email{dakinimallam@gmail.com}
\and Salim Abdou Daoura \at ML Collective, \email{salim.adaoura@gmail.com}}
%
%
\maketitle

\abstract{Multi-organ segmentation is often challenged by partially annotated datasets and domain shifts across different imaging sources. To address these limitations, we propose a two-stage learning framework that efficiently leverages partial supervision.
In the first stage, the model learns from available annotations to produce accurate segmentations of annotated organs, establishing robust feature representations. In the second stage, we introduce learnable organ prototypes and a Sinkhorn-triplet loss to enforce organ-wise feature consistency across datasets. This encourages latent embeddings of the same organ to remain close, while increasing separation between different organs, even when annotations are missing.
Our approach achieves performance comparable to state-of-the-art methods on the BTCV dataset, while remaining computationally efficient.
By explicitly aligning feature distributions rather than relying solely on pseudo-labels, the framework effectively mitigates domain shift, making it particularly suitable for medical image segmentation tasks with limited annotation resources.
 }


\section{Introduction}

Automatic organ segmentation plays a central role in various applications, such as diagnostic imaging, radiotherapy planning, and pathology analysis. Deep Learning methods have emerged as the dominant paradigm for this task, but they traditionally require a large number of fully annotated scans. Since manual annotation is inherently time-consuming, often only partially annotated datasets are available. \\

As a result, many existing datasets provide only partial annotations focused on a single target organ. Training separate models for each dataset is computationally inefficient, while approaches that share an encoder but use task-specific decoders fail to fully exploit inter-organ relationships. \\

\begin{figure}[t]
\sidecaption[t]
\includegraphics[width=0.66\textwidth]
{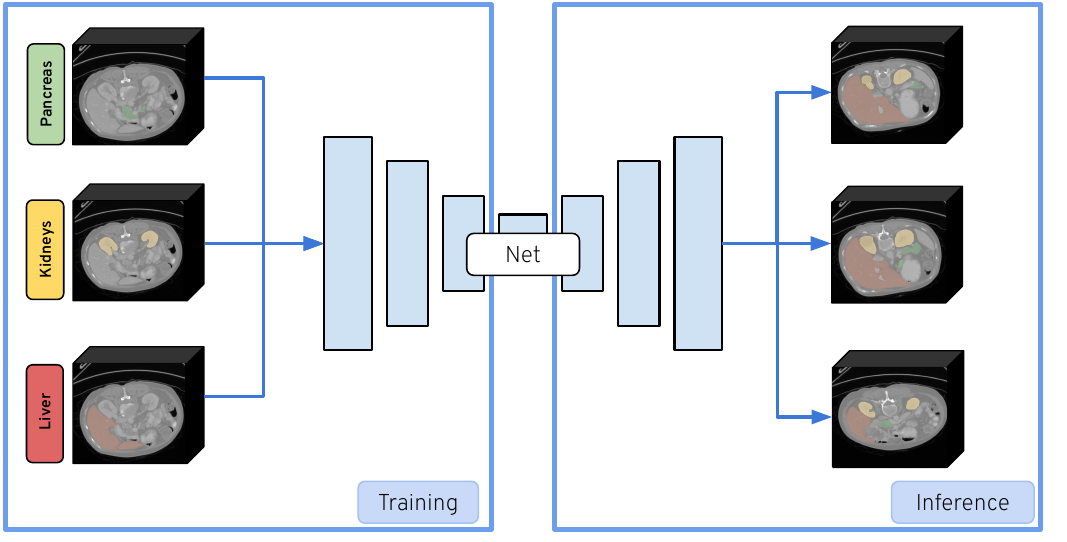} 
\caption{Our training setting: From partially annotated datasets, we aim to learn a single unified model that will produce predictions for all organs.}
\label{fig1}
\end{figure}

To address this limitation, we propose a two-stage training framework that learns from partially annotated datasets to build a single unified model, as illustrated in Figure \ref{fig1}. In the first stage, learning is performed using only the available ground truth of the partial datasets. We employ adapted segmentation losses that encourage agreement between ground truth and prediction. This stage establishes reliable feature representations.

In the second stage, we employ an optimal transport-based triplet loss that encourages organ-wise feature consistency across datasets. This regularization helps the network learn coherent representations for both annotated and unannotated domains, mitigating domain shift.

For instance, when the liver is annotated in one dataset but unannotated in another, our model regularizes their feature distributions to remain close in the latent space, ensuring semantic consistency. Simultaneously, it enforces separation from the distributions of other organs to improve inter-class discriminability. We compare features predicted for unlabeled organs against a set of learned prototype features derived from the ground truth annotations. The proposed Sinkhorn-triplet loss replaces pointwise distances with optimal transport–based divergences, aligning entire feature distributions rather than individual samples.

Our contributions are summarized as follows:
\begin{itemize}
    \item We propose a novel application of the Sinkhorn-triplet loss to partially supervised medical image segmentation. By replacing pointwise distances in the triplet formulation with Sinkhorn divergences between feature distributions, our method enforces distribution-level alignment between predicted and ground truth feature spaces.
    
    \item We design a training strategy where pseudo-labels are used solely to determine anchor class membership, rather than as direct supervision signals. This prevents the direct propagation of pseudo-label errors and promotes robust feature alignment through optimal transport–based regularization.
    
    \item Our approach enhances feature consistency and cross-dataset generalization, enabling accurate segmentation of unannotated organs by aligning features across domains at the distributional level.
\end{itemize}

\section{Related Work}

\subsection{Partially Supervised Segmentation}
Early efforts to train unified models from partially annotated datasets often focused on modifying the loss function. \cite{gonzalez2018multi} introduced a masked Dice loss that leverages only available ground truth labels during backpropagation. \cite{schutera2022methods} further refined this idea by incorporating a class-asymmetric objective function (CAL) that penalizes incorrect predictions under the assumption of mutual exclusivity. While these approaches effectively ignore missing annotations, they do not exploit the pseudo-labels generated by the model itself, thus overlooking potentially valuable semantic information that could enhance segmentation performance. \\

An alternative research direction focused on architectural design.
DoDNet \cite{zhang2021dodnet} and OmniSeg \cite{deng2021omni} implemented a dynamic segmentation head, where specific convolutional kernels are assigned to each task for the segmentation of a specific organ and tumors. The kernels are adaptively generated by a task controller, conditioned on both the input image and the assigned task. This mechanism enables the network to adapt its decoding path to the segmentation of specific organs or lesions, offering greater flexibility and computational efficiency compared to training multiple independent models or using multi-head decoders.
However, these methods still rely solely on ground truth supervision and require multiple forward passes at inference time to segment all organs, limiting their scalability.\\

More recent approaches make use of the pseudo-labels generated by the model via a two-stage training strategy. In COSST \cite{liu2024cosst}, the network first learns from available annotation, then generates pseudo-labels for missing organs. The quality of these pseudo-labels is assessed by outlier detection in the latent space. Finally, these reliable pseudo-labels are iteratively used for finetuning. 
PSSNET \cite{liu2024many} also adopts a two-stage learning process where in the second stage, the authors retrain the segmentation model using an adversarial network as a means to reduce the discrepancy between annotated and unannotated domains.
Although these strategies filter pseudo-labels, they still risk propagating noisy supervision when the filtering mechanism fails, potentially destabilizing training and degrading generalization.
More recently, \cite{jiang2025labeledtounlabeled}
showed promising results by introducing a labeled-to-unlabeled distribution alignment framework.

\subsection{Deep Metric Learning}
Deep Metric Learning has been used in various applications such as Person RE-Identification \cite{varior2016siamese}, few shot learning \cite{vinyals2016matching} and face recognition \cite{schroff2015facenet}.
Its goal is to train an embedding model to measure the similarity between data points. Images are mapped into feature vectors through a deep neural network encoder. A loss function is tailored to minimize the distance between similar vectors, and maximize the distance between vectors of different classes. Some standard losses used are the contrastive loss \cite{hadsell2006contrastive} or the triplet loss \cite{schroff2015facenet}. The triplet loss operates on triplets of samples : an anchor, a positive point (from the same class) and a negative point (from a different class). The distance of the positive pair is pulled to be smaller than the negative pair. 
It creates an embedding space where features of the same class are close while the inter-class distance is high. 
In our framework, we adapt this principle to medical image segmentation by introducing a modified triplet objective that enforces organ-wise feature separability across datasets.

\subsection{Optimal Transport}
Optimal transport (OT) determines the cost of transforming one distribution into another. It captures geometric correspondences between distributions, making it particularly effective for comparing structured feature spaces.
OT has been used in a variety of computer vision tasks: domain adaptation \cite{ding2023cross}, multitask learning \cite{janati2019wasserstein}, feature matching \cite{sarlin2020superglue} or generative model \cite{arjovsky2017wasserstein}. 
Recent advances have also leveraged OT for self-supervised representation alignment \cite{gorade2025otcxr} and topology-aware segmentation losses based on OT distances between persistent diagrams \cite{demir2023topology}.

In this work, we exploit an OT-based distance to compare feature distributions across organs.
Unlike pointwise metrics, OT accounts for the underlying geometry of the embedding space, providing a more robust measure of feature separability and alignment across partially annotated datasets.

\section{Method}

\begin{figure*}[t]
\centering
\includegraphics[width=0.9\textwidth]{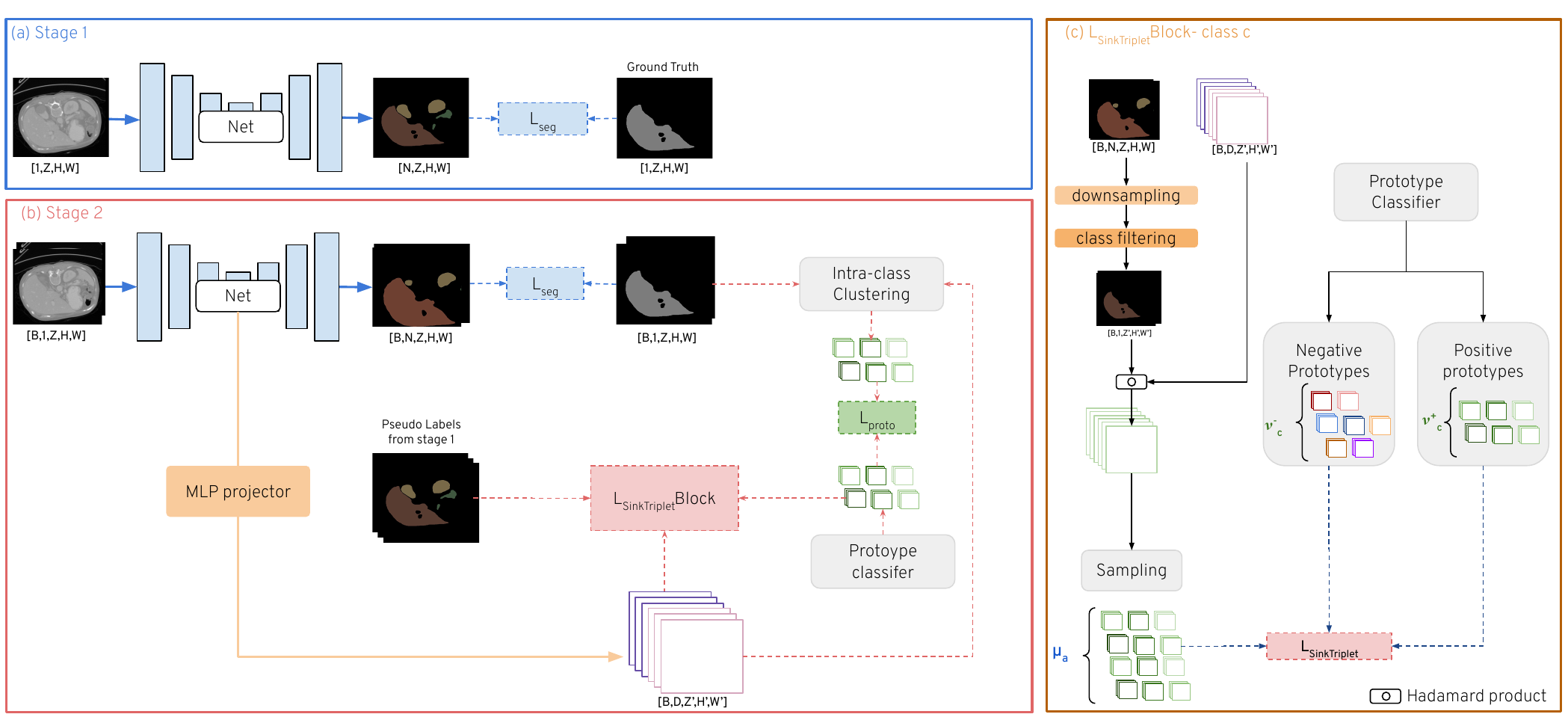} 
\caption{Overview of the proposed two-stage framework. (a) In Stage 1, the model is trained using only the available ground truth annotations available in partially labeled datasets. At the end of stage 1, pseudo-labels are generated and filtered. (b) In Stage 2, training is regularized with the Sinkhorn-triplet loss that enforces inter-class feature alignement. A prototype classifier learns each class prototypes (c) The Sinkhorn-triplet loss is computed. Anchor distribution is sampled based on the pseudo-labels from stage 1 while the positive and negative distributions are derived from the prototype classifier. }
\label{fig2}
\end{figure*}

In this section, we formalize our two-stage training framework. Figure \ref{fig2} provides an overview of our proposed method.

\subsection{Problem Formulation}
We consider $N$ partially labeled datasets 
\[
D_{pl} = \{ D_1, \dots, D_N \} \quad \text{s.t.} \quad D_i = \{x,y\}, \ \forall i \in [1,\dots,N].
\]
For each dataset $D_i$, only one organ $C_i$ is annotated in the ground truth mask $y$.  
Our goal is to learn a single unified model 
$
\mathcal{F}_{\theta}: [1, Z, H, W] \mapsto [N, Z, H, W]$ from the collection of partially labeled data $D_{pl}$.

\subsection{Stage 1: Supervised Learning with Ground Truth}
The first stage of our framework consists in learning exclusively from the available annotations.
The objective is to maximize the agreement between the predictions of the labeled organs and the corresponding ground truth.
Given a prediction $\tilde{y} = F_\theta(x)$, the segmentation objective is defined as:
\begin{equation}
\mathcal{L}_{seg} = \mathcal{L}_{Dice}(\tilde{y}_t, y) + \mathcal{L}_{CE}(\tilde{y}_t, y)
\end{equation}
Here, $\tilde{y}_t$ denotes the predicted probabilities restricted to the labeled organ.
$\mathcal{L}_{Dice}$ and $\mathcal{L}_{CE}$ correspond to the standard Dice loss and cross-entropy loss, respectively.
After stage 1, pseudo-labels are generated for the training set, and a confidence threshold $\tau$ is applied, based on the softmax predictions, to retain only the most reliable predictions.

\subsection{Stage 2: Feature consistency with the Sinkhorn-triplet loss}
Approaches that directly leverage pseudo-labels are often prone to noise propagation, as unreliable predictions may be treated as ground truth and destabilize training.
Instead, we use pseudo-labels solely to infer class membership in unlabeled regions, without using them as explicit supervision targets. This allows us to define positive and negative pairs while relying only on prototypes derived from ground-truth annotations to guide semantic alignment across datasets.

\subsubsection{Multimodal prototype learning}

In the second stage of training, we define a set of learnable prototypes that serve as compact and multimodal representations of each organ class.
Each class $c$ is associated with $K$ prototypes $\mathcal{P}c = \{p_{c,1}, \dots, p_{c,K}\}$, initialized randomly and optimized jointly with the network parameters.
During training, feature embeddings $F$ are extracted from a mid-level layer of the network and projected into a lower-dimensional space. For each class $c$, we compute local centroids $\mathcal{C}_c$ from the features corresponding to that class using an online clustering step.
The prototypes are then updated to minimize their distance to these centroids, ensuring that they capture the underlying modes of the class distribution.
To prevent excessive fragmentation of the feature space, we introduce a variance regularization term that enforces intra-class compactness, encouraging prototypes of the same class to remain close to their mean representation. 

The prototype learning loss is formulated as :

\begin{equation}
    L_{proto} = \frac{1}{C}\sum_{c = 1}^{C}   \left\lVert   \mathcal{P}_c - \mathcal{C}_c \right\lVert ^2+\frac{\lambda_{compact}}{C}\sum_{c = 1}^{C} Var(\mathcal{P}_c)
\end{equation}

\subsubsection{The triplet loss}

The classical triplet loss operates on triplets of samples 
$(x_a, x^+, x^-)$ corresponding respectively to an anchor, 
a positive example from the same class, and a negative example from a different class.
It is defined for a triplet as:
\begin{equation}
    \mathcal{L}_{triplet} =
    \max\big(d(x_a, x^+) - d(x_a, x^-) + m, 0\big),
\end{equation}
where $d$ is typically the Euclidean distance, and $m$ is a fixed margin encouraging inter-class separation.

At each iteration, the stage-1 pseudo-label mask $\tilde{Y}_{batch}$ selects feature vectors assigned to class $c$. Each selected vector $x_a$ serves as an anchor, while the positive and negative examples are drawn from the learned prototype sets:
\begin{itemize}
    \item  $x^+$ is the closest prototype in $\mathcal{P}_c$, and
    \item $x^-$ is the closest prototype from another class $c' \neq c$.
\end{itemize}
This ensures that the alignment process is driven by ground-truth derived class prototypes while pseudo-labels are only used to guide anchor selection in unlabeled regions.

\subsubsection{The Sinkhorn-triplet loss}
Following \cite{dou2022optimaltriplet}, we generalize the triplet formulation to operate on distributions rather than individual embeddings, replacing the pointwise distance by the Sinkhorn divergence~\cite{feydy2019interpolating}.

We define each anchor, positive, and negative example as a discrete probability measure supported in the feature space $\mathbb{R}^D$.  
The anchor distribution $\mu_a$ is constructed from the pseudo-labeled region predicted as class $c$. Instead of using all voxels, we randomly sample $K_a$ feature vectors $f_i^a$ from this region and assign them uniform weights:
\begin{equation}
\mu_a = \frac{1}{K_a}\sum_{i=1}^{K_a}\delta_{f_i^a}, \quad f_i^a \in \{\,f \mid \tilde{Y}_{batch}(f) = c\,\}.
\end{equation}
This sampling reduces the impact of noisy or redundant pseudo-labeled points.

The positive distribution $\nu_{c}^{+}$ corresponds to the ground-truth prototype distribution of the same class :
\begin{equation}
\nu_{c}^{+} = \frac{1}{K_c}\sum_{k=1}^{K_c}\delta_{p_{c,k}}, \quad p_{c,k} \in \mathcal{P}_c,
\end{equation}
and the negative distribution $\nu^{-}$ is defined as the union of prototype distributions from all other classes:
\begin{equation}
\nu_{c}^{-} = \frac{1}{\sum_{c' \neq c}K_{c'}} \sum_{c' \neq c} \sum_{k=1}^{K_{c'}} \delta_{p_{c',k}}, \quad p_{c',k} \in \mathcal{P}_{c'}.
\end{equation}

The Sinkhorn divergence between two empirical distributions $\mu$ and $\nu$ is defined as:
\begin{equation}
S_{\epsilon,p}(\mu, \nu)
= OT_{\epsilon,p}(\mu, \nu)
- \frac{1}{2}OT_{\epsilon,p}(\mu, \mu)
- \frac{1}{2}OT_{\epsilon,p}(\nu, \nu),
\end{equation}
where $OT_{\epsilon,p}(\mu,\nu)$ is the optimal transport cost regularized by entropy:
\begin{equation}
OT_{\epsilon,p}(\mu, \nu)
= \min_{\pi \in \Pi(\mu,\nu)}
\langle C_p, \pi \rangle + \epsilon \, \mathrm{KL}(\pi \,||\, \mu \otimes \nu),
\end{equation}
with cost $C_p(x,y) = \frac{1}{p}\|x - y\|^p$ and regularization parameter $\epsilon$.

The resulting Sinkhorn-triplet loss is then expressed as:
\begin{equation}
\mathcal{L}_{SinkTriplet} =
\max\Big(
S_{\epsilon,p}(\mu_a, \nu_{c}^+) -
S_{\epsilon,p}(\mu_a, \nu_{c}^-) + m, \, 0
\Big),
\end{equation}

This formulation enforces that the predicted feature distribution of a pseudo-labeled region aligns with its corresponding class prototype distribution while remaining distant from other classes in the Wasserstein space.

The overall objective function used in the second training stage combines segmentation and alignment objectives:
\begin{equation}
\mathcal{L} = \mathcal{L}_{seg} + \lambda_{triplet} * \mathcal{L}_{SinkTriplet} + \lambda_{proto} * \mathcal{L}_{proto} 
\end{equation}
where $\lambda_{triplet}$  and $\lambda_{proto}$ are weighting coefficients.


\section{Experiments}
\subsection{Datasets}
We evaluated our approach using four widely adopted organ-labeled datasets:
Kits19 \cite{heller2021kits19} containing kidney and kidney tumor annotations, Lits19 \cite{bilic2023liver} focusing on liver segmentation MSD \cite{antonelli2022msd} Pancreas and Spleen datasets.
Tumor masks were merged with their corresponding organ masks. In addition to the held-out test sets of each dataset, we employed the BTCV dataset \cite{landman2015btcv} as an external benchmark to assess cross-dataset generalizability.
Table \ref{table:datasets} in the appendix summarizes the different dataset splits. \\

\subsection{Experiments Setup}

\subsubsection{Implementation details}
All experiments were conducted with the nnUNet framework \cite{isensee2021nnu}, chosen for its strong and robust performance in biomedical image segmentation. 
For preprocessing, all CT scans are resampled to 2×0.80×0.80 mm, clipped to [-1024,1024] Hounsfield Units (HU), and rescaled to [0,1].
The patch size used by nnUNet is 96 x 160 x 160. 

For stage 2, we used the Geomloss library \cite{feydy2019interpolating} to compute the Sinkhorn divergence. 
Our feature projector is a small MLP composed of two Conv3d layers with kernel size of 1, separated by a batch normalization and ReLU activation layers. 
A complete list of the hyperparameters used is available in the appendix.

\subsubsection{Evaluation metrics}
We report three widely used metrics: the Dice Similarity Coefficient (DSC), 95th Hausdorff distance (HD95) and the Average Surface Distance (ASD).
DSC measures spatial overlap between the prediction mask and the ground truth. 
HD95 is the 95th percentile of the maximum distances between the boundary points in the prediction and the ground truth. 
ASD measures the average distance between the predicted segmentation and the ground truth along the surface of the segmented object.

\subsection{Comparison with the State-of-the-art}
We evaluated our method against seven state-of-the-art (SOTA) partial-label segmentation methods, with all baseline results reported from \cite{liu2024cosst}. For a fair comparison, our method uses the same architecture and training strategy.
SOTA methods include those using individual networks trained on each partial dataset (denoted Multi-Net), those using only ground-truth supervision (TAL \cite{fang2020tal} and ME \cite{shi2021marginal}), those using pseudo-label supervision (COSST \cite{liu2024cosst}, PLT \cite{gonzalez2018multi}, and Co-training \cite{huang2020cotraining}), and the conditioned network DoDNet \cite{zhang2021dodnet}.
We exclude kidney results due to a preprocessing discrepancy: the baseline methods treat left and right kidneys as two separate labels, whereas our method treats them as a single class. A summary of the comparison is presented in Table \ref{table:task1_performance_reduced}. \\

\begin{table*}[t]
\centering
\setlength{\tabcolsep}{6pt}
    \renewcommand{\arraystretch}{1} 
\caption{Performance comparison (DSC, \%, higher is better; HD95, mm, lower is better; ASD, mm, lower is better) of partially labeled segmentation methods. P: held-out testing sets of Kits19, MSD and Lits19. F: BTCV. Our method achieves the best DSC on the partial datasets while staying comparable to SOTA methods on BTCV dataset.}
\begin{tabular}{llllllllll}
\hline\noalign{\smallskip}
\multicolumn{1}{c}{\textbf{Methods}} & \multicolumn{3}{c}{\textbf{Spleen (P)}} & \multicolumn{3}{c}{\textbf{Pancreas (P)}} & \multicolumn{3}{c}{\textbf{Liver (P)}} \\ 
& DSC & HD95 & ASD & DSC & HD95 & ASD & DSC & HD95 & ASD \\
\noalign{\smallskip}\svhline\noalign{\smallskip}
Multi-Nets    & 91.59 & 26.15 & 3.72 & 77.21 & 6.88 & 1.40 & 94.81 & 7.54 & 1.28 \\
TAL      & 93.72 & \textcolor{blue}{2.97} & 0.82 & 76.36 & 6.37 & 1.43 & 94.86 & 6.10 & 1.31 \\
ME        & 94.09 & 2.38 & \textcolor{blue}{0.77} & 75.99 & 6.01 & \textcolor{red}{0.77} & 94.70 & 7.45 & 1.90 \\
PLT      & 93.41 & 34.85 & 3.97 & 77.69 & 6.82 & 1.32 & \textcolor{blue}{95.24} & 6.13 & \textcolor{blue}{1.10} \\
Co-train  & 93.09 & 42.06 & 5.24 & 77.50 & 6.22 & 1.21 & \textcolor{blue}{95.24} & \textcolor{red}{5.10} & \textcolor{red}{0.96} \\
DoDNet    & 93.75 & 15.44 & 2.23 & \textcolor{blue}{79.09} & \textcolor{blue}{5.84} & \textcolor{blue}{1.17} & 93.65 & 11.26 & 2.03 \\
COSST  & \textcolor{blue}{94.40} & \textcolor{red}{1.45} & \textcolor{red}{0.54} & 77.98 & \textcolor{red}{5.82} & 1.29 & 95.22 & \textcolor{blue}{5.78} & 1.18 \\
\noalign{\smallskip}\svhline\noalign{\smallskip}
Ours & \textcolor{red}{95.52} & 15.03 & 2.72 & \textcolor{red}{83.83} & 9.44 & 1.83 & \textcolor{red}{96.00}& 12.30 & 2.16 \\

\noalign{\smallskip}\hline\noalign{\smallskip}
\end{tabular}

\vspace{1em}

\begin{tabular}{llllllllll}
\hline\noalign{\smallskip}
\multicolumn{1}{c}{\textbf{Methods}} & \multicolumn{3}{c}{\textbf{Spleen (F)}} & \multicolumn{3}{c}{\textbf{Pancreas (F)}} & \multicolumn{3}{c}{\textbf{Liver (F)}} \\
& DSC & HD95 & ASD & DSC & HD95 & ASD & DSC & HD95 & ASD \\
\noalign{\smallskip}\svhline\noalign{\smallskip}
Multi-Nets    & 87.72 & 29.36 & 6.21 & 73.11 & 22.06 & 4.14 & 94.80 & 7.19 & 1.18 \\
TAL       & 90.02 & \textcolor{blue}{12.15} & \textcolor{blue}{2.28} & 73.86 & 7.77 & 1.76 & \textcolor{blue}{95.99} & 8.01 & 1.05 \\
ME        & \textcolor{blue}{90.58} & 14.25 & 2.91 & 74.87 & 21.77 & 3.65 & 95.96 & 10.15 & 1.60 \\
PLT       & 88.00 & 55.87 & 8.17 & 75.38 & 11.24 & 2.13 & 95.82 & \textcolor{red}{3.65} & \textcolor{red}{0.62} \\
Co-train  & 88.53 & 52.43 & 8.82 & 75.93 & 10.81 & 1.90 & 95.82 & 5.18 & 0.77 \\
DoDNet   & 89.94 & 22.01 & 4.50 & 76.51 & \textcolor{blue}{7.34} & 1.61 & 94.45 & 20.44 & 3.58 \\
COSST & \textcolor{red}{92.34} & \textcolor{red}{4.13} & \textcolor{red}{1.25} & \textcolor{blue}{77.17} & \textcolor{red}{5.90} & \textcolor{blue}{1.36} & \textcolor{red}{96.28} & \textcolor{blue}{3.86} & \textcolor{blue}{0.71} \\
\noalign{\smallskip}\svhline\noalign{\smallskip}
Ours & 90.03 & 20.77 & 4.73 & \textcolor{red}{81.27} & 7.73 & \textcolor{red}{1.26} & 94.32 & 15.82 & 2.44 \\

\noalign{\smallskip}\hline\noalign{\smallskip}
\end{tabular}

 \label{table:task1_performance_reduced}
\end{table*}

On the held-out test sets corresponding to the partial datasets, our proposed method demonstrates superior segmentation accuracy in terms of the Dice Score across all evaluated organs. Notably, an improvement of over 4\% in DSC is observed for the pancreas compared to all competing methods, highlighting our model's effectiveness on challenging structures. Furthermore, when considering boundary localization metrics, our approach achieves a lower HD95 for the spleen compared to methods relying on multiple networks (Multi-Nets, PLT, Co-train). This suggests that our unified model not only performs well but also offers greater computational efficiency compared to these approaches. 
However, we observe relatively high HD95 values across several organs, particularly for the liver, which is the largest organ evaluated. This suggests a potential limitation where the model prioritizes volumetric overlap leading to high DSC, especially for large organs, over precise boundary delineation. This behavior is likely exacerbated by our proposed regularization loss, which primarily focuses on ensuring feature distribution coherence rather than explicitly penalizing fine-grained contour inaccuracies. \\

On the BTCV dataset, our method achieves performance comparable to state-of-the-art techniques across most organs and metrics, while demonstrating notably superior results for pancreas segmentation. More importantly, our model exhibits improved robustness to domain shift compared to baseline methods. For example, the Multi-Nets approach experiences a performance degradation of 4.1\% in DSC for the pancreas when moving from the partial test sets to BTCV (77.21\% to 73.11\%). In contrast, our method shows a significantly smaller drop of only 2.56\% for the same organ under the same domain shift. This indicates that the feature consistency enforced by our regularization mechanism enhances the model's ability to generalize to unseen data distributions.

\subsection{Ablation Studies}

We conducted an ablation study to measure the improvements induced by our different method components. 

\subsubsection{Quantitative results}

\begin{table*}[t]
\centering
\caption{Ablation study on our method components on the BTCV dataset. The addition of the second training stage improves the model segmentation capibility, with all losses. The Sinkhorn-triplet loss gets the highest improvements overall.}

\begin{tabular}{lllllllllllll}
\hline\noalign{\smallskip}
\multicolumn{1}{c}{\textbf{Methods}} & \multicolumn{3}{c}{\textbf{Spleen }} & \multicolumn{3}{c}{\textbf{Pancreas }} & \multicolumn{3}{c}{\textbf{Liver }} & \multicolumn{3}{c}{\textbf{Kidney }}\\
& DSC & HD95 & ASD & DSC & HD95 & ASD & DSC & HD95 & ASD & DSC & HD95 & ASD \\
\noalign{\smallskip}\svhline\noalign{\smallskip}
Baseline & 89.61 & 21.59 & \textcolor{red}{4.33} & 80.87 & 9.20 & 1.44 & 94.06 & 17.70 & 2.81 & 81.55 & 27.32 & 5.13  \\
w/ $\mathcal{L}_{triplet}$ & \textcolor{red}{90.03} & 28.17 & 5.38 & 81.03 & 8.54 & 1.35 & 94.30 & 16.95 & 2.66 & 82.67 & 26.45 & 4.92  \\
w/ $\mathcal{L}_{Sinkhorn}$ & 89.93 & \textcolor{red}{20.76} & 5.06 & 81.05 & 8.56 & 1.39 & 94.08 & \textcolor{red}{15.65} & 2.63 & \textcolor{red}{83.82} & \textcolor{red}{24.26} & \textcolor{red}{4.75}  \\
w/ $\mathcal{L}_{sinkTriplet}$ & \textcolor{red}{90.03} & 20.77 & 4.73 & \textcolor{red}{81.27} & \textcolor{red}{7.73} & \textcolor{red}{1.26} & \textcolor{red}{94.32} & 15.82 & \textcolor{red}{2.44} & 83.50 & 26.67 & 4.83 \\

\noalign{\smallskip}\hline\noalign{\smallskip}
\end{tabular}
 \label{table:sota_performance}
\end{table*}

Table \ref{table:sota_performance}
summarizes the quantitative results on  the BTCV dataset. Baseline is the model at the end of stage 1, the results of stage 2 are compared with a Sinkhorn divergence loss, the standard triplet loss and the proposed Sinkhorn-triplet loss. 

Both the Triplet loss and the Sinkhorn loss, when added individually, demonstrate improvements in average DSC compared to the baseline. Specifically, $\mathcal{L}_{Sinkhorn}$ yields noticeable gains in Kidney DSC ie. 81.55\% vs 83.82\%, along with improvement in HD95 for Spleen and Liver. $\mathcal{L}_{Triplet}$ also shows modest DSC improvements across most organs but less consistently improves boundary metrics.

 $\mathcal{L}_{SinkTriplet}$ achieves the highest DSC scores across most organs, surpassing both the baseline and the models trained with individual losses. Notably, Liver DSC increases from 94.06\% (Baseline) to 94.32\%, and Kidney DSC improves from 81.55\% to 83.50\%. This strongly suggests that simultaneously optimizing for feature separability and intra-class compactness leads to the most robust volumetric segmentation performance when generalizing to unseen data.

\subsubsection{Separability in the embedding space}
\begin{figure}[t]
    \centering
    \begin{tikzpicture}
        \begin{axis}[
            colormap name=viridis,
            xtick={0,1,2,3,4},
            ytick={0,1,2,3,4},
            xticklabels={Background, Pancreas, Kidney, Liver, Spleen},
            yticklabels={Background, Pancreas, Kidney, Liver, Spleen},
            y dir=reverse,
            xticklabel style={rotate=30, anchor=north east},
            enlargelimits=false,
            axis on top,
            point meta min=0, point meta max=1,
            nodes near coords={\pgfmathprintnumber[fixed, precision=2]{\pgfplotspointmeta}},
            nodes near coords style={font=\tiny, color=white, anchor=center},
            width=0.40\textwidth,
            height=0.40\textwidth
        ]
        \addplot[matrix plot*, mesh/cols=5, point meta=explicit] coordinates {
            (0,0) [0.0000] (1,0) [0.5238] (2,0) [0.5473] (3,0) [0.2399] (4,0) [0.3994]
            (0,1) [0.5238] (1,1) [0.0000] (2,1) [0.5128] (3,1) [0.5979] (4,1) [0.7996]
            (0,2) [0.5473] (1,2) [0.5128] (2,2) [0.0000] (3,2) [0.7986] (4,2) [0.6580]
            (0,3) [0.2399] (1,3) [0.5979] (2,3) [0.7986] (3,3) [0.0000] (4,3) [0.2067]
            (0,4) [0.3994] (1,4) [0.7996] (2,4) [0.6580] (3,4) [0.2067] (4,4) [0.0000]
        };
        \end{axis}
    \end{tikzpicture}
    \hspace{0.0\textwidth} 
    \begin{tikzpicture}
        \begin{axis}[
            colorbar,
            colormap name=viridis,
            xtick={0,1,2,3,4},
            ytick={0,1,2,3,4},
            xticklabels={Background, Pancreas, Kidney, Liver, Spleen},
            yticklabels={Background, Pancreas, Kidney, Liver, Spleen},
            y dir=reverse,
            xticklabel style={rotate=30, anchor=north east},
            enlargelimits=false,
            axis on top,
            point meta min=0, point meta max=1,
            nodes near coords={\pgfmathprintnumber[fixed, precision=2]{\pgfplotspointmeta}},
            nodes near coords style={font=\tiny, color=white, anchor=center},
            width=0.40\textwidth,
            height=0.40\textwidth
        ]
        \addplot[matrix plot*, mesh/cols=5, point meta=explicit] coordinates {
            (0,0) [0.0000] (1,0) [0.5695] (2,0) [0.5852] (3,0) [0.2541] (4,0) [0.4047]
            (0,1) [0.5695] (1,1) [0.0000] (2,1) [0.5933] (3,1) [0.7836] (4,1) [0.9205]
            (0,2) [0.5852] (1,2) [0.5933] (2,2) [0.0000] (3,2) [0.8296] (4,2) [0.7113]
            (0,3) [0.2541] (1,3) [0.7836] (2,3) [0.8296] (3,3) [0.0000] (4,3) [0.1259]
            (0,4) [0.4047] (1,4) [0.9205] (2,4) [0.7113] (3,4) [0.1259] (4,4) [0.0000]
        };
        \end{axis}
    \end{tikzpicture}
    \caption{Heatmaps of inter-class distances after stage 1 (left) and after stage 2 (right) on the BTCV dataset. Overall, stage 2 achieves higher inter-class distances, yielding better separability.}
    \label{fig:distance_matrices}
\end{figure}
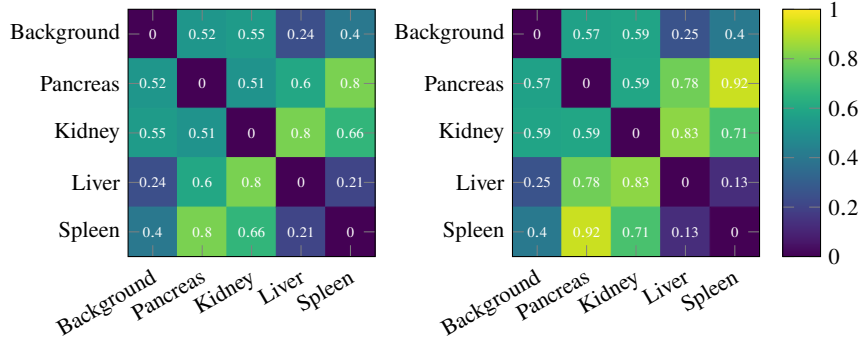

To quantify the quality of the feature space, we measured the Calinski-Harabasz (CH) Index, which evaluates both intra-class compactness and inter-class separability. The results for Stage 1 and Stage 2 are presented in Table \ref{tab:ch_index}. We also computed the distances between class centroids in the embedding space. As shown in Figure \ref{fig:distance_matrices}, the results highlight consistently higher distances between the majority of classes in the second stage. Notably, the centroid distance between the Spleen and Pancreas, two anatomically adjacent organs, improved from 0.80 to 0.92. This suggests that the proposed Sinkhorn-Triplet loss effectively enhances inter-class separability, yielding more accurate segmentation results.

\begin{table}[h]
    \centering  
    \caption{Comparison of Calinski-Harabasz index score in embedding space between Stage 1 and Stage 2 (Higher is better). Stage 2 demonstrates better intra-class compactness and inter-class separability.}
    \begin{tabular}{lc}
        \hline\noalign{\smallskip}
        \textbf{Model} & \textbf{Calinski-Harabasz Index}  \\
        \hline\noalign{\smallskip}
        Baseline & 743.44  \\
        \hline\noalign{\smallskip}
        w/ $\mathcal{L}_{sinkTriplet}$ & \textcolor{red}{958.20}  \\
        \noalign{\smallskip}\hline\noalign{\smallskip}
    \end{tabular}
     
     \label{tab:ch_index}
\end{table}

\subsubsection{Qualitative Results}
\begin{figure*}[t]
\centering
\includegraphics[width=0.8\textwidth]{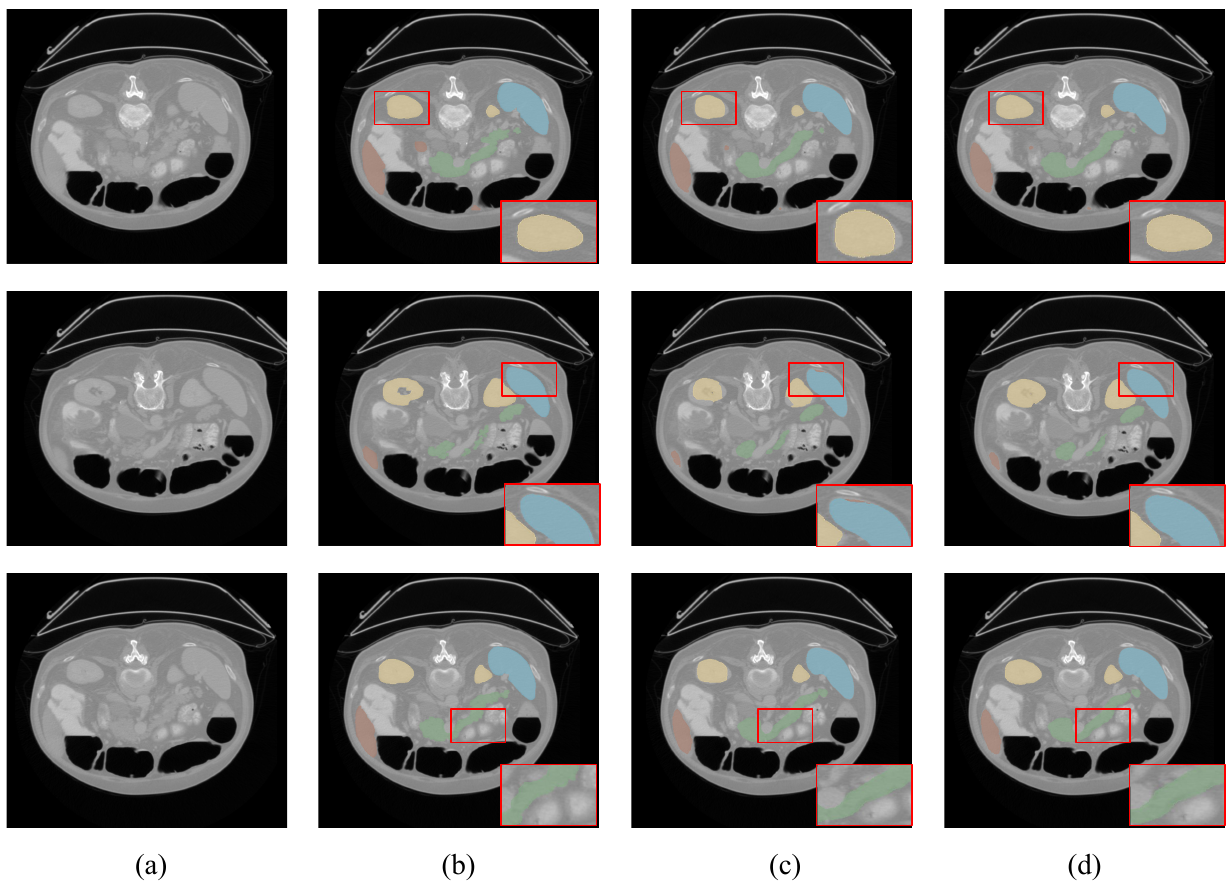} 
\caption{Visualization of segmentation results after Stage 1 and Stage 2 of our framework. (a) Original scan, (b) Ground truth annotation, (c) Stage 1 result, (d) Stage 2 result. Our framework enables more accurate voxel predictions and corrects errors from Stage 1.}
\label{fig3}
\end{figure*}

Figure \ref{fig3} provides a qualitative comparison between the segmentation outputs of the baseline model and our proposed solution. The visualizations highlight more accurate segmentations achieved by our method after Stage 2, with particularly evident improvements for structures like the kidneys. We attribute this enhanced segmentation quality to the better-defined class clusters in the learned feature space.

\subsubsection{Sensitivity to hyperparameters}
\begin{figure}[t]
    \centering

    \begin{minipage}[t]{0.45\textwidth}
        \centering
        \begin{tikzpicture}
            \begin{axis}[
                xlabel={$margin$},
                ylabel={$Dice\ Score$},
                width  = \linewidth,
                height = 4cm,
                xmin=0, xmax=1.6,
                ymin=86.9, ymax=87.16,
                xtick={0.1, 0.5, 1, 1.2, 1.5},
                ytick={86.9,87,87.1,87.2},
                legend pos=south east
            ]
            \addplot[smooth,mark=*,bblue] plot coordinates {
                (0.1,87.11) (0.5,87.15) (1,87.09) (1.2,86.93) (1.5,86.93)
            };
            \end{axis}
        \end{tikzpicture}
        \label{fig:ablation-margin}
    \end{minipage}
    \hfill
    \begin{minipage}[t]{0.45\textwidth}
        \centering
        \begin{tikzpicture}
            \begin{axis}[
                xlabel={$\varepsilon$ (blur)},
                ylabel={$Dice\ Score$},
                width  = \linewidth,
                height = 4cm,
                xmode=log,
                xmin=0.01, xmax=100,
                ymin=86.85, ymax=87.15,
                xtick={0.01,0.1,1,10},
                ytick={86.9,87.0,87.1},
                legend pos=south west
            ]
            \addplot[smooth,mark=*,bblue] plot coordinates {
                (0.03,87.09) (0.1,87.04) (1,86.91) (10,86.86)
            };
            \end{axis}
        \end{tikzpicture}
        \label{fig:ablation-epsilon}
    \end{minipage}

    \caption{Ablation study showing the effect of margin (left) and  Sinkhorn blur parameter $\varepsilon$ (right) on mean Dice score. Our method is overall robust to these hyperparamters, with slightly better result for a margin of 0.5 and small blur $\varepsilon$ values.}
    \label{fig:ablation-both}
\end{figure}
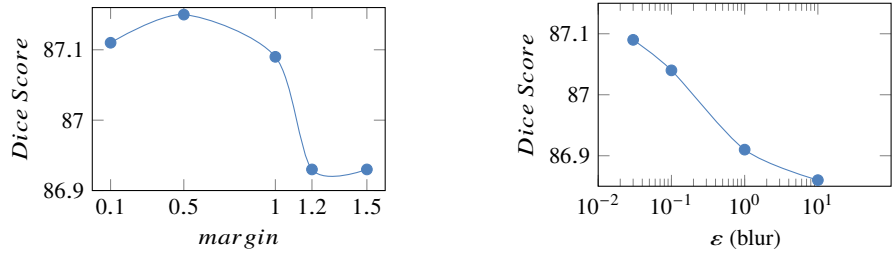

We investigated the impact of the margin hyperparameter on the overall segmentation performance. The margin controls the desired separation between positive and negative distributions in the learned embedding space. We trained our full model with varying margin values, keeping all other hyperparameters constant, and evaluated the mean Dice score on the BTCV validation set.

As shown in Figure \ref{fig:ablation-both} (left), the model's performance exhibits relative robustness to the choice of margin, particularly within the range of 0.1 to 1.0. The optimal mean Dice score of 87.15\% was achieved with a margin of 0.5. Margins greater than or equal to 1.2 led to a more noticeable drop in performance, suggesting that enforcing an overly large separation might hinder the optimization process.
We also made several experiments with varying blur value $\varepsilon$ which controls the entropy regularization in the Sinkhorn divergence calculation. Results shown in Figure \ref{fig:ablation-both} (right) indicate that the highest performance is achieved with the smallest tested blur value $\varepsilon$=0.03. Performance consistently decreases slightly as the blur increases, suggesting that a more precise OT approximation is marginally beneficial. However, the overall performance variation across the tested range is minimal (less than 0.25\%), demonstrating that the method is relatively robust to this hyperparameter.

\section{Limitations}
We acknowledge some limitations that open avenues for future research.
The proposed Sinkhorn-triplet loss is designed to minimize the divergence between feature distributions. While effective for semantic consistency, this global alignment objective does not explicitly penalize local boundary errors. Future work could investigate integrating boundary-aware constraints, such as active contour losses, directly into the optimal transport formulation. Another Limitation is that our two-stage approach relies on the initial supervised model to generate pseudo-labels that define the anchor distributions for Stage 2. Exploring frameworks that can dynamically refine pseudo-labels without a fixed multi-stage pipeline is a promising direction to mitigate this dependency.

\section{Conclusion}
We introduced a novel framework for multi-organ segmentation under partial supervision, designed to effectively handle datasets with incomplete annotations.
Our method reduces the representation gap between labeled and unlabeled domains through feature-level alignment.
We believe that this approach provides a promising direction towards improving domain generalization and robustness in medical image segmentation tasks.

        

    

    

            
    

            
    
    


\ethics{Competing Interests}{
The authors have no conflicts of interest to declare that are relevant to the content of this chapter.}

\eject

\section*{Appendix}
\addcontentsline{toc}{section}{Appendix}

\subsection*{Datasets splits used for experiments}
For training and validation, we used the four public partial datasets Kits19 \cite{heller2021kits19}, Lits19 \cite{bilic2023liver} and Spleen and Pancreas from MSD \cite{antonelli2022msd}. The BTCV dataset is only used for test, to assess our method generalization capabilities. The different splits used are showed on Table \ref{table:datasets}.
\begin{table}[h]
\setlength{\tabcolsep}{8pt} 
    \renewcommand{\arraystretch}{1} 

\centering
\caption{Dataset splits used in our experiments.}
\begin{tabular}{llll}
\hline\noalign{\smallskip}
\textbf{Dataset} & \textbf{Training} & \textbf{Validation} & \textbf{Test} \\
\hline\noalign{\smallskip}
KiTS19           & 80       & 20        & 105 \\
\hline\noalign{\smallskip}
LiTS19           & 52       & 13         & 60  \\
\hline\noalign{\smallskip}
Pancreas (MSD)   & 112      & 28         & 140 \\
\hline\noalign{\smallskip}
Spleen (MSD)    & 16        & 4         &  21 \\
\hline\noalign{\smallskip}
BTCV             & –        & –          & 30  \\
\noalign{\smallskip}\hline\noalign{\smallskip}
\end{tabular}
\label{table:datasets}
\end{table}

\subsection*{Experiments setting}
experiments were conducted using a NVIDIA A100 40Go. Table \ref{stage1_hparams}
and \ref{stage2_hparams} gather the hyperparameters used for stage 1 and stage 2 respectively. 
The model with the best dice performance during stage 1 was used to initialize stage 2. 

\begin{table}[h]
\centering

\begin{minipage}[t]{0.45\textwidth}
\centering
\caption{Stage 1 hyperparameters.}
\begin{tabular}{l c}
\hline\noalign{\smallskip}
\textbf{Hyperparameter} & \textbf{Value} \\
\hline\noalign{\smallskip}
Optimizer & SGD \\
Learning rate ($lr$) & 1e-3 \\
Number of epochs & 1000 \\
Momentum & 0.99 \\
Confidence threshold ($\tau$) & 0.9 \\
\noalign{\smallskip}\hline\noalign{\smallskip}
\end{tabular}
\label{stage1_hparams}
\end{minipage}
\hfill
\begin{minipage}[t]{0.45\textwidth}
\centering
\caption{Stage 2 hyperparameters.}
\begin{tabular}{l c}
\hline\noalign{\smallskip}
\textbf{Hyperparameter} & \textbf{Value} \\
\hline\noalign{\smallskip}
Learning rate ($lr$) & 5e-5 \\
Learning rate (MLP) & 1e-4 \\
Learning rate (prototype parameters) & 1e-4 \\
Number of epochs & 100 \\
Momentum & 0.99 \\
Number of prototypes per class ($K$) & 5 \\
Triplet margin ($m$) & 0.5 \\
\hline\noalign{\smallskip}
Prototype loss weight ($\lambda_{\text{proto}}$) & 1 \\
Compactness weight ($\lambda_{\text{compact}}$) & 0.1 \\
Triplet loss weight ($\lambda_{\text{triplet}}$) & 1 \\
\hline\noalign{\smallskip}
Sinkhorn divergence blur ($\epsilon$) & 0.03 \\
Sinkhorn scaling factor & 0.9 \\
\noalign{\smallskip}\hline\noalign{\smallskip}
\end{tabular}
\label{stage2_hparams}
\end{minipage}
\end{table}

%
%


%
%
%

\end{document}